\documentclass[aps,article,nofootinbib,groupedaddress]{revtex4}
\usepackage{graphicx}
\usepackage{amsmath,amssymb}
\usepackage{hyperref}
\usepackage[usenames]{color}

\hypersetup{
    colorlinks=true,
    linkcolor=red,
    citecolor=blue,
}

\def\be{\begin{equation}}
\def\ee{\end{equation}}
\def\ba{\begin{eqnarray}}
\def\ea{\end{eqnarray}}

\begin{document}

\title{
Soliton Creation with a Twist
}

\date{\today}

\author{Tanmay Vachaspati
}

\affiliation{Department of Physics, Arizona State University,
Tempe, AZ 85287.}

\begin{abstract}
We consider soliton creation when there are ``twist'' degrees of 
freedom present in the model in addition to those that make up the
soliton. Specifically we consider a deformed O(3) sigma model in
1+1 dimensions, which reduces to the sine-Gordon model in the
zero twist sector. We study the scattering of two or more breather 
solutions as a function of twist, and find soliton creation for a
range of parameters. We speculate on the application of these ideas, 
in particular on the possible role of magnetic helicity, to the production 
of magnetic monopoles, and the violation of baryon number in nuclear
scattering experiments.
\end{abstract}

\maketitle


Field theories are known to have two sectors -- small excitations
called ``particles'', and large excitations called ``solitons''.
Perturbative quantum field theory deals with particles and their 
interactions. Solitons are generally studied as classical solutions, 
unrelated to the particle description. An outstanding open problem 
in quantum field theory is to describe solitons as an assembly of 
particles. A closely related problem is the central focus of this
paper -- how can we build solitons by scattering particles?

The relation between solitons and particles is explicitly known in the
sine-Gordon model \cite{Mandelstam:1975hb} where the soliton
operator is given in terms of an infinite number of particle
operators. Thus to build solitons from particles in the sine-Gordon 
model requires assembly of an infinite number of particles and
is impossible. This ties in with the complete integrability 
of the sine-Gordon model wherein soliton number is preserved.
Therefore it is necessary to consider models that are not completely 
integrable. However integrable models are amenable to a lot of
analysis, and it seems unwise to abandon them altogether. 
We would like to take advantage of the insights that are provided
by integrable models and to use them in a non-integrable setting. 
Hence we aim to start with a completely integrable model, modify it 
suitably to eliminate complete integrability while still preserving 
as many features as possible, and then study soliton creation in the 
non-integrable model. 

There can of course be many modifications of a completely integrable 
model. Indeed, the $\lambda \phi^4$ model in 1+1 dimensions studied in 
\cite{Dutta:2008jt,Romanczukiewicz:2010eg,Demidov:2011eu} can also be 
viewed as a deformation of the sine-Gordon model. Here we would like to study
a different kind of modification; one that may be useful in generalization 
to higher spatial dimensions where we would like to study the creation of
vortices and magnetic monopoles. The modification is inspired by recent 
work within the electroweak model \cite{Copi:2008he,Chu:2011tx}: the
annihilation of ``twisted'' magnetic monopoles leads to the production
of magnetic fields that inherit the twist in the form of magnetic helicity. 
We will describe this observation in greater detail in 
Sec.~\ref{generalizations}. For the time being we note that magnetic 
monopoles can carry a relative twist, and so we would like to introduce a 
similar twist degree of freedom in the 1+1 dimensional sine-Gordon model, 
thus also breaking complete integrability. Then the kinks of the sine-Gordon 
model will be able to carry relative twist, and solitons may be created in 
scattering that involves the twist degree of freedom. The hope is that 
twisted initial conditions are better suited to the creation of solitons. 
In our toy model in Sec.~\ref{O3model} we will see that twist is 
{\it essential} for the creation of solitons because untwisted initial 
conditions lie in a completely integrable sub-space of the model.

A different view of the soliton creation problem also suggests that 
twist can play a useful role. When two particles scatter, it is advantageous 
if they spend a long time together, so that other particles also have time 
to interact and create a multi-particle state such as the soliton-antisoliton 
pair. In our case, we will take the initial state to be a sequence of 
sine-Gordon breathers -- the closest we can come to a sequence of particles
in classical theory. When two in-phase 
breathers scatter in the sine-Gordon model, they pass through each other 
but with a {\it negative} time delay {\it i.e.} a time advance. This is 
because of an attractive force between the breathers. If the breathers 
are twisted, however, the force can be repulsive, and the scattering can 
lead to a time delay. In Sec.~\ref{twobreathers} we will study the scattering 
of two breathers as a function of twist. The numerical results confirm that 
twist can lead to a positive time delay in the scattering

The paper is organized as follows. In Sec.~\ref{O3model}
we describe our toy model, and various solutions in it.
In Sec.~\ref{twobreathers} we study the scattering of two breather 
solutions of the model and show that twisted breathers lead
to a time delay in the scattering. As a special case we find
the time delay in the scattering of two sine-Gordon breathers.
In Sec.~\ref{manybreathers} we study the scattering of two 
sequences (``trains'') of twisted breathers by numerically evolving 
the equations of motion. Here we find examples of soliton creation.
We discuss generalizations to higher spacetime dimensions, and
to vortices and monopoles in Sec.~\ref{generalizations}. The production
of magnetic monopoles is intimately related to baryon number violation
via electroweak sphaleron production. We discuss this connection in
Sec.~\ref{generalizations}, where we also conclude.
In Appendix~\ref{appa} we provide some notes related 
to the numerical evolution of the constrained O(3) system.

\section{Deformed O(3) model}
\label{O3model}

With the motivation described above, we now move on to describe a model 
in 1+1 dimensions that contains kinks and also allows for twist.
We start with the O(3) sigma model given by the action
\begin{equation}
S_\sigma = \int d^2 x \frac{1}{2} (\partial_\mu {\hat n})^2
\end{equation}
where ${\hat n} = {\hat n}(t,x)$ satisfies the constraint
\begin{equation}
{\hat n}^2 =1
\end{equation}
This model is known to be completely integrable \cite{Zamolodchikov:1978xm}
and we will need to spoil this feature. So we consider
\begin{equation}
S = \int d^2 x \left [
               \frac{1}{2} (\partial_\mu {\hat n})^2
              -\frac{1}{2} (1-n_3^2) \right ]
\end{equation}
where $n_3$ is the third (or $z-$) component of ${\hat n}$.
The additional potential term spoils the O(3) invariance and
we now only have $Z_2$ symmetry under $n_3 \to -n_3$ and
O(2) symmetry under rotations about the z-axis. The true
vacua are
\begin{equation}
{\hat n} = \pm (0,0,1)
\end{equation}
The model has kink solutions that interpolate between the 
two vacua and there is a one parameter family of such kinks.

One way to proceed is by introducing a Lagrange multiplier,
$\xi$, to enforce the constraint ${\vec n}^2 =1$
\begin{equation}
L = \frac{1}{2} (\partial_\mu {\vec n} )^2 
      - \frac{\xi}{2} ( 1 - {\vec n}^2 )
      - \frac{1}{2} [ 1 - n_3^2 ]
\end{equation}
This leads to the equation of motion
\begin{equation}
\square {\vec n} + (\partial_\mu {\vec n})^2 {\vec n}
      - n_3 ( {\hat e}_3 - n_3 {\vec n} ) =0, \ \ 
{\vec n}^2 = 1
\label{nevolve}
\end{equation}

To gain some intuition, we consider an explicit solution to the constraint 
equation 
\begin{equation}
{\hat n} = (\sin\theta \cos\phi, \sin\theta \sin\phi, \cos\theta )
\label{nthetaphi}
\end{equation}
Then
\begin{equation}
L = \frac{1}{2} (\partial_\mu \theta )^2
    + \frac{1}{2} \sin^2\theta (\partial_\mu \phi )^2
     -\frac{1}{2} \sin^2 \theta
\end{equation}
and the equations of motion are
\begin{eqnarray}
\square \alpha + \sin\alpha (1 - (\partial_\mu \phi)^2 ) &=& 0
   \label{thetaeq} 
    \\
\partial_\mu ( \sin^2\theta \partial^\mu \phi ) &=& 0
   \label{phieq}
\end{eqnarray}
where $\alpha \equiv 2\theta$.

If $\phi$ is constant, then
\begin{equation}
\square \alpha = - \sin\alpha
\end{equation}
which is the equation of motion for the sine-Gordon model, and all 
solutions of the sine-Gordon model are also solutions of our deformed 
O(3) model. Hence our model contains embedded spaces given by 
$\phi={\rm constant}$ where the dynamics is that of the sine-Gordon
model\footnote{In quantum theory fluctuations in 
the $\phi$ direction will contribute to the dynamics of $\theta$.}. Thus 
our model contains a one 
parameter family of sine-Gordon models, one per circle of longitude;
the other degree of freedom, $\phi$, will be referred to as the 
``twist''.

The kink solution in our model is exactly the same as in the sine-Gordon
model
\begin{equation}
\theta_k = 2 \tan^{-1}[e^x],\ \ \phi={\rm constant}
\label{kinksolution}
\end{equation}
The energy in the fields is given by
\begin{equation}
E = \int dx H(t,x) = \int dx \left [ 
   \frac{{\dot\theta}^2}{2} +
   \frac{{\theta '}^2}{2} +
   \sin^2\theta\frac{{\dot\phi}^2}{2} +
   \sin^2\theta\frac{{\phi'}^2}{2} + 
   \frac{1}{2}\sin^2\theta \right ]
\label{hamiltonian}
\end{equation}
and for the kink evaluates to
\begin{equation}
E_k = 2 
\label{kinkenergy}
\end{equation}
Boosted breather solutions can also be borrowed from the sine-Gordon model
\begin{equation}
\theta_b (t,x;x_0,v)=2\tan^{-1}\left[ 
        \frac{\eta \sin(\omega T)}{\cosh(\eta\omega X)} \right ], \ \ 
\phi = {\rm constant}
\label{thetab}
\end{equation}
where $\omega \in [0,1]$ is a parameter -- the oscillation frequency of the
breather -- and 
\begin{equation}
\eta = \omega^{-1}\sqrt{1-\omega^2} \ .
\end{equation}
The boosted coordinates $(T,X)$ are given by
\begin{equation}
T = \gamma [t-v(x-x_0)],\ X =\gamma [(x-x_0) - vt]
\end{equation}
where $x_0$ is the location of the breather at $t=0$, 
$v$ is the speed of the breather, and $\gamma = 1/\sqrt{1-v^2}$ is the 
Lorentz boost factor. 

The energy of a boosted breather is
\begin{equation}
E_b = 4\gamma \sqrt{1-\omega^2}
\label{breatherenergy}
\end{equation}

Additional solutions can be obtained if $\partial^\mu \phi = k^\mu$, a constant 
two vector, then again
\begin{equation}
\square_y \alpha = - \sin\alpha
\end{equation}
where the D'Alembertian operator now contains derivatives with respect to 
$y^\mu \equiv \sqrt{1-k^2} x^\mu$. The $\phi$ equation is satisfied provided
$k^\mu \partial_\mu \theta =0$. This can be arranged by choosing
$\theta = \theta (p_\mu x^\mu)$ with $k\cdot p =0$. For example, if we take 
$k^\mu = (\kappa, 0)$, the solution is ``spinning'' in field space. In 
this case, $\theta$ can be any static solution of the sine-Gordon model, 
in particular, it can be the sine-Gordon kink. This solution can be pictured 
on the two-sphere as lying on a line of longitude that spins with angular 
frequency $\kappa$.

The feature that our model has sine-Gordon subspaces is advantageous becase
breathers are non-dissipative solutions in the model. Thus we can formulate
non-dissipative initial conditions by writing them in terms of
a sequence of breathers. In the quantized sine-Gordon model, the smallest
breather corresponds to a particle, and hence, a sequence of classical
breathers is the closest we can come classically to a sequence of 
quantum particles \cite{Dashen:1975hd}.

\section{Scattering of two breathers}
\label{twobreathers}

We first study the scattering of two breathers and find the time delay
in the scattering. In the untwisted case, this is well-defined because
the breathers pass through each other. In the twisted case, the breathers
get deformed on interaction. However, we can still calculate the time
delay in the propagation of the center of energy. We will elaborate on
this below.

The initial conditions corresponding to two incoming breathers
with twist $\xi$ are
\begin{eqnarray}
\theta(t=0,x) &=& \theta_b(t=0,x;-x_0,+v) + \theta_b(t=0,x;+x_0,-v) 
\label{2bic}\\
\phi(t=0,x) &=& \pi \xi ~  {\rm tanh}(x/w) \\
{\dot \theta}(t=0,x) &=& {\dot \theta}_b(t=0,x;-x_0,v) + 
                           {\dot \theta}_b(t=0,x;+x_0,-v) 
\label{eq:dottheta}
\\
{\dot \phi}(t=0,x) &=& 0
\label{eq:dotphi}
\end{eqnarray}
where $\theta_b (t=0,x;x_0,v)$ is Eq.~(\ref{thetab}) evaluated at $t=0$ and 
${\dot \theta}_b(t=0,x;x_0,v)$ is obtained by differentiating Eq.~(\ref{thetab})
with respect to time and then evaluating at $t=0$. The expressions are
sufficiently messy that we don't display them. The $\phi$ field is
taken to have a ${\rm tanh}$ profile with a width $w$ that is much smaller
than $x_0$. Since $\theta (0,x) \simeq 0$ near $x=0$, the precise choice of $w$ 
is not significant.

One subtlety arises since, in spherical coordinates, $\theta \in [0,\pi]$ 
and $\phi \in [0,2\pi]$,
whereas in Eq.~(\ref{2bic}) $\theta$ can become negative. This will not
be of consequence, however, since we will always work with the vector
${\hat n}$ using Eq.~(\ref{nthetaphi}), and $\partial_t {\hat n}$. 
In the construction of ${\hat n}$ and $\partial_t{\hat n}$,
negative $\theta$ yields the same vector as positive 
$\theta$ but with $\phi \to \phi+\pi$. 

The domain of the twist parameter $\xi$ is determined by noting that 
the total change in $\phi$ is $2\pi \xi$. Hence it is sufficient to 
take $\xi \in [0,0.5]$, since $\xi=0.5$ corresponds to 
$\phi(x=-\infty)=-\pi/2$ and $\phi(x=+\infty)=+\pi/2$ {\it i.e.}
the dynamics occurs on the great circle in the $yz-$plane, where 
it is given by the sine-Gordon model. For $\xi=0$, the incoming
breathers are in phase; for $\xi=0.5$, the incoming breathers
are out of phase.

The numerical algorithm is described in Appendix~\ref{appa}.
To find the time delay, we evaluate the position of the center of 
energy on half the space
\begin{equation}
x_{\rm CE}(t) = \frac{1}{E/2} \int_0^\infty dx x H(t,x)
\end{equation}
where $H$ is the Hamiltonian density (Eq.~(\ref{hamiltonian})) 
and $E$ is the total energy. The time delay after some long 
time $T$ is given by
\begin{equation}
\delta t = T - \frac{x_{\rm CE}(T) }{v}
\label{deltat}
\end{equation}
This formula assumes that all the energy in the final state is 
propagating with speed $v$ and so the time delay makes strict sense 
only if the incoming breathers pass through each other after 
interacting and retain their identity. This will only happen for 
$\xi=0,~0.5$ when the dynamics is sine-Gordon. For other values 
of $\xi$, the scattering causes some radiation, though most of 
the energy still resides in a breather-like lump and the definition
in Eq.~(\ref{deltat}) can be used as a measure of the
time delay.

\begin{figure}
  \includegraphics[height=0.45\textwidth,angle=0]{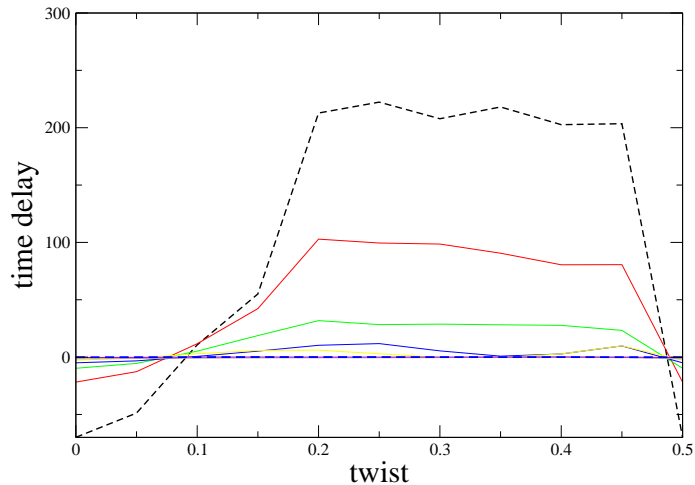}
  \caption{Time delay vs. twist for $\omega=0.86$ and $v$ ranging from 
0.1 (dashed curve with larger amplitude) to 0.9 (dashed curve with smaller 
amplitude). The time delay for $v=0.9$ is at the $10^{-2}$ level though
the dependence on twist is similar to that of the other curves.
}
\label{twisttimedelayom}
\end{figure}

\begin{figure}
  \includegraphics[height=0.45\textwidth,angle=0]{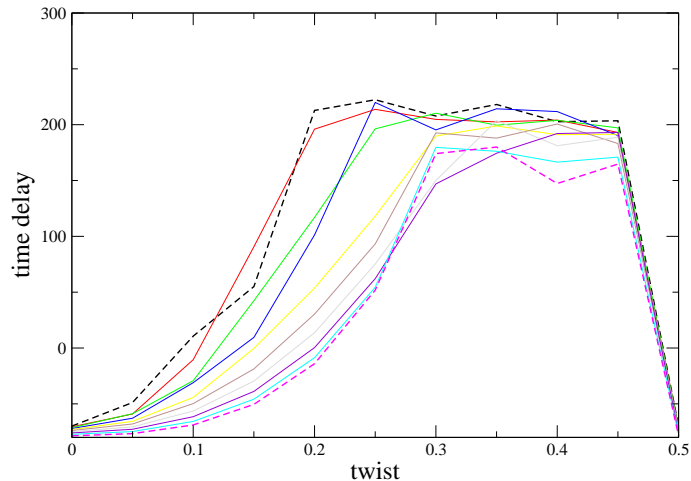}
  \caption{Time delay vs. twist for $v=0.1$ and $\omega$ ranging from $0.86$ 
(dashed curve with larger amplitude) to 0.95 (dashed curve with smaller amplitude).
}
\label{twisttimedelayv}
\end{figure}

In Fig.~\ref{twisttimedelayom} we plot the time delay versus twist 
for fixed $\omega=0.86$ and for incoming velocity ranging from 0.1 
to 0.9. The plot shows that there is an equal time advance for the
in-phase ($\xi=0$) and out-of-phase ($\xi=0.5$) sine-Gordon 
breathers\footnote{Here we disagree with Ref.~\cite{Nishida:2009}
where a time delay is found for out-of-phase scattering on the
basis of a proposed collective coordinate method.}.
The plot also shows that twist can cause a time delay and the
effect is largest at low velocities. Similarly in 
Fig.~\ref{twisttimedelayv} we see the role of twist on the
time delay for a variety of breathers of different frequency,
$\omega$, but with fixed $v=0.1$. The time delay is largest
for the low frequency ({\it i.e.} large amplitude) breathers.

\section{Scattering of many breathers}
\label{manybreathers}

Now we consider the scattering of two trains of $N$ identical breathers.
The total incoming energy has to be larger than the kink-antikink energy.
With the kink and breather energies in Eqs.~(\ref{kinkenergy}) and
(\ref{breatherenergy}), we get the condition
\begin{equation}
8N\gamma \sqrt{1-\omega^2} > 4
\end{equation}
which translates into
\begin{equation}
N > \frac{1}{2}\sqrt{\frac{1-v^2}{1-\omega^2}}
\label{Nbound}
\end{equation}
This condition is simply an energy requirement. 

\begin{figure}
  \includegraphics[height=0.45\textwidth,angle=0]{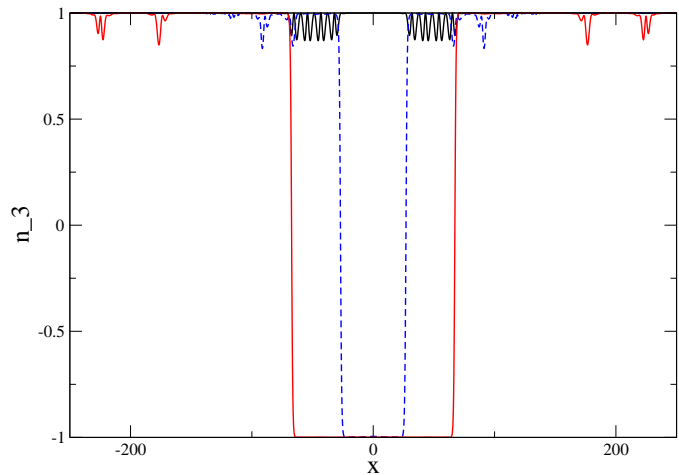}
  \caption{Three snapshots of the $z-$component of ${\hat n}$ for
the run with $v=0.5$, $\xi=0.1$ and the remaining parameters as
in Eq.~(\ref{fixedparameters}). The initial breather state is the 
black solid curve. The final state is the solid (red) curve with 
the kink and the antikink interpolating between $n_3=+1$ and
$n_3=-1$.  An intermediate state is also shown (dashed blue curve).
}
\label{kinkcreation}
\end{figure}

The initial condition is a generalization of 
Eqs.~(\ref{2bic})-(\ref{eq:dotphi})
\begin{eqnarray}
\theta(t=0,x) &=& \sum_{j=1}^{N}
    \left [ \theta_b(t=0,x;-x_j,+v) + \theta_b(t=0,x;+x_j,-v) \right ]
\label{Nbic}\\
\phi(t=0,x) &=& \pi \xi ~  {\rm tanh}(x/w) \\
{\dot \theta}(t=0,x) &=& \sum_{j=1}^N
    \left [{\dot \theta}_b(t=0,x;-x_j,v) + 
                           {\dot \theta}_b(t=0,x;+x_j,-v) \right ] \\
{\dot \phi}(t=0,x) &=& 0 
\label{eq:Ndotphi}
\end{eqnarray}
with
\begin{equation}
x_j = x_0 + (j-1) a
\end{equation}
where $x_0$ is the location of the innermost breather and $a$ is the
spacing between the breathers in a train. An example of the initial
condition is shown in Fig.~\ref{kinkcreation}.

\begin{figure}
  \includegraphics[height=0.45\textwidth,angle=0]{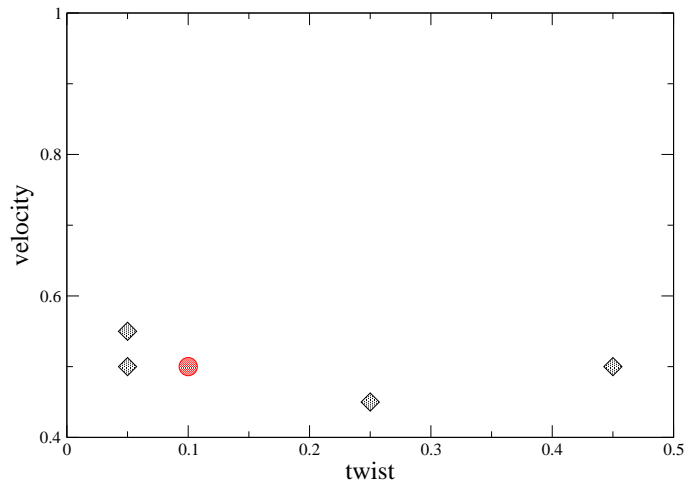}
  \caption{
Diamonds indicate points in the velocity-twist plane where we have 
found soliton creation to occur with parameters as in 
Eq.~(\ref{fixedparameters}). The point denoted by the filled circle 
corresponds to the parameters used in the successful kink creation
shown in Fig.~\ref{kinkcreation}.
Our numerical strategy scans in parameter space in steps of 0.05 
in twist in the interval [0,0.5], and steps of 0.05 in velocity
in the interval [0.4,0.95]. A more refined scan would fill in
some of the parameter space though we expect a fractal distribution
of parameters which successfully produce kinks 
\cite{Dutta:2008jt,Romanczukiewicz:2010eg} 
}
\label{velvstwist}
\end{figure}

With the initial conditions in terms of $\theta$ and $\phi$, we 
construct the vector field ${\hat n}(0,x)$ and its time derivative, 
$\partial_t{\hat n}(0,x)$. We then evolve ${\hat n}$ using 
Eq.~(\ref{nevolve}), as discussed in greater detail in
Appendix~\ref{appa}. We hold the following parameters fixed
in our numerical runs
\begin{equation}
N=4,\ \ \omega=0.95,\ \ x_0=32,\ \ a=\frac{4}{\eta\omega\gamma}=11.1,
\ \ w=\frac{2}{\eta \omega \gamma}=5.55,
\label{fixedparameters}
\end{equation}
and $v \in [0,1]$ and $\xi \in [0,0.5]$ are scanned over. Most
runs do not yield soliton creation but some runs do. An example 
of soliton creation occurs for $v=0.5$ and $\xi =0.1$, and is 
shown in Fig.~\ref{kinkcreation}. In 
Fig.~\ref{velvstwist} we mark the points in parameter space 
where we have found soliton creation. Note that non-zero twist 
is essential for soliton creation in this model.

It is worth pointing out that the outcome is sensitive to the
choice of our fixed parameters in Eq.~(\ref{fixedparameters}).
Dependence on $N$ and the spacing $a$ is expected; but we also
find dependence on $x_0$, suggesting that the phase of the 
breather trains on collision may be important in the outcome.
This would be consistent with the chaotic nature of kink
production in $\lambda \phi^4$ observed in 
\cite{Dutta:2008jt,Romanczukiewicz:2010eg}. 
In future work we are planning a more extensive scan of 
parameter space and to examine the chaotic nature of the
scattering more closely \cite{ongoingwork}.

\section{Generalizations and Conclusions}
\label{generalizations}

The deformed O(3) sigma model suggests many generalizations that
all seem worthwhile to explore. 

The first generalization is to stay in 1+1 dimensions and to extend 
the model to an $N-$component vector. (This $N$ is not to be confused 
with the number of breathers in the previous section.) It may be 
possible to make some analytic advances by considering the kink 
creation problem in the large $N$ approximation. 
Other generalizations are to complexify the vector ${\hat n}$
and consider the CP(N) model, and to also include gauge fields.

The sigma model framework can be used to generalize to vortex
and monopole formation, by simply choosing different potentials.
We have already studied kink creation in the O(3) model with 
the potential
\begin{equation}
V({\hat n}) = \frac{1}{2} [1-n_3^2]
\end{equation}
To study vortex formation in 2+1 dimensions, it would be
natural to choose
\begin{equation}
V({\hat n}) = \frac{1}{2} [1-(n_1^2+n_2^2)]
\end{equation}
Now the true vacua lie around the equator of an $S^2$, and
the vortex solutions correspond to either the northern or
the southern hemisphere, and the center of the vortex will
be either at the north or south poles. These two solutions
are related by the $Z_2$ symmetry under $n_3 \to -n_3$. 
In this model, we do not have a ``twist'' degree of freedom 
as in the kink case. Recall that in the kink case, we had 
a 1-parameter family of kink solutions labeled by the constant 
value of the azimuthal field $\phi$ as in Eq.~(\ref{kinksolution}).
Similarly we need a one parameter family of vortex solutions
and this can be obtained by extending the model to O(4).
Then the vacua are still given by circles $n_1^2+n_2^2=1$ but 
there is a one parameter family of possible vortex solutions
labeled by the angle in the $(n_3,n_4)$ plane. 

A point to remember about vortices is that the energy of a global 
vortex-antivortex pair grows logarithmically with separation.
So their creation is necessarily transient, unless the model
is gauged. A similar comment applies to the creation of magnetic
monopoles. 

To study monopole formation in 3+1 dimensions in our O(3) model
we would choose the potential,
\begin{equation}
V({\hat n}) = \frac{1}{2} [1-(n_1^2+n_2^2+n_3^2)]
\label{monopolepotential}
\end{equation}
except that this vanishes due to the constraint. The O(3) model
does have the correct topology ($\pi_2(S^2)=\mathbb{Z}$) for magnetic 
monopoles, but the monopoles are singular because there is no
escape from the vacuum manifold. If we choose the potential
in Eq.~(\ref{monopolepotential}) but in an O(4) model, we get
a situation similar to the vortex discussed above in the O(3) 
model: there are two non-singular monopole solutions corresponding
to the two ``hemispheres'' in $S^3$. To include a twist similar
to the one for kinks, we should go to at least an O(5) model, 
where the one parameter family of monopole solutions is labeled 
by the angle in the $(n_4,n_5)$ plane. However, monopoles can
be relatively twisted even in the O(3) model and we now discuss
this feature in greater detail. 

Let us assume that a monopole is located at $z=+a$ and its asymptotic
vector field ${\hat n}$ is given by Eq.~(\ref{nthetaphi}) with 
$\theta = \theta_+$, where $\theta_+$ is the spherical angle
measured from the $+z-$axis but with the origin of the coordinate
system at $z=+a$. An antimonopole is now placed at $z=-a$ and its 
asymptotic field is given by
\begin{equation}
{\hat n} = (\sin{\bar \theta} \cos{\bar \phi}, 
            \sin{\bar \theta} \sin{\bar \phi}, 
            \cos{\bar \theta} )
\end{equation}
with ${\bar \theta} = \pi - \theta_-$ where $\theta_-$ is the
spherical angle measured from the $+z-$axis but with origin
at $z=-a$. The azimuthal angle ${\bar \phi} = \phi + \alpha$
and $\alpha$ plays the role of the twist in the O(3) model.

In Ref.~\cite{Taubes:1982ie} Taubes used a Morse theory construction
in the gauged O(3) model and proved the existence of a static solution 
consisting of a monopole and an antimonopole ($m{\bar m}$) separated by 
some distance. This is surprising because one expects the Coulombic 
$-1/r^2$ attractive force between the $m{\bar m}$) to bring 
them together, after which the two can annihilate. In Taubes' solution, 
the $m{\bar m}$ are kept apart by a repulsive force 
provided by the ``twist''. The result is an unstable solution since 
the monopoles can untwist, come together, and annihilate. The solution 
itself is quite complicated but has been found numerically in
Refs.~\cite{Ruber:1985,Kleihaus:1999sx}.

A construction similar to the one used by Taubes exists in the electroweak 
model and the resulting $m{\bar m}$ solution is called the ``sphaleron'' 
\cite{Dashen:1974ck,Manton:1983nd} (for a review see \cite{Manton:2004tk}). 
The form of the sphaleron solution is spherically symmetric for a particular 
value of coupling constants and is known explicitly up to two simple profile
functions given by ordinary second order differential equations. The solution 
has been studied numerically for a large range of coupling constants
\cite{Kleihaus:1991ks,Kunz:1989,Yaffe:1989ms}. 

A particularly convenient feature of the Morse theory construction, 
is that it provides a path from the $m{\bar m}$ solution to the vacuum. 
Thus the sphaleron decay path describes the annihilation 
of a twisted $m{\bar m}$ pair. As shown in 
Refs.~\cite{Copi:2008he,Chu:2011tx} the decay products of the 
annihilation include electromagnetic fields with non-zero magnetic 
helicity\footnote{In the SU(2) limit of the electroweak model there
are no photons in the model. In this limit, sphaleron decay yields 8 Higgs 
particles and 42 W particles \cite{Hellmund:1991ub}.}, which is defined as
\begin{equation}
{\cal H} = \int d^3x {\bf A}\cdot {\bf B}
\end{equation}
The magnetic helicity, which measures the twisting and linking of 
magnetic field lines, is inherited from the initial twist in the $m{\bar m}$ 
pair that make up the sphaleron. The sphaleron decay analysis also shows
that, after a short period of time, the evolution of the electroweak
fields is such that magnetic helicity is conserved. 

By a time reversal of the sphaleron decay process, we conclude that
the {\it convergence} of electromagnetic helicity can possibly lead to the 
creation of sphalerons, and hence also $m{\bar m}$ with a relative 
twist. The twist may also provide the necessary repulsive force between 
monopoles and antimonoples so as to separate them and to prevent them 
from re-annihilating.  

The usefulness of magnetic helicity may be conjectured based on another
line of reasoning. In highly conducting plasmas it is known that magnetic 
helicity is a conserved quantity. The derivation goes as follows:
\begin{eqnarray}
\frac{d{\cal H}}{dt} &=& 
\int d^3x  \left [~ \partial_t {\bf A}\cdot {\bf B}
                +  {\bf A} \cdot \partial_t{\bf B}~ \right ]
\nonumber \\
&=& 
\int d^3x  \left [ ({\bf E}+{\bf \nabla}A^0 )\cdot {\bf B}
             - {\bf A} \cdot {\bf \nabla}\times {\bf E} \right ]
\nonumber \\
&=&
\int d^3x  \left [ 2 {\bf E}\cdot {\bf B}
             + ({\bf \nabla} A^0) \cdot {\bf B} \right ] 
\nonumber \\
        &=& \int d^3 x \left [ 2{\bf E}\cdot {\bf B} 
              - {A^0} {\bf \nabla}\cdot {\bf B} \right ] 
\nonumber \\
        &=& 2 \int d^3 x \left [ \left (\frac{\bf j}{\sigma} 
              + {\bf v}\times {\bf B} \right )\cdot {\bf B} ] \right ]
\nonumber \\
 &\to& 0
\end{eqnarray}
In the last two steps we have used Ohm's law as applied to a plasma 
element moving with velocity ${\bf v}$. The electrical conductivity is 
assumed to be infinite, $\sigma \to \infty$, though there are arguments
in support of helicity conservation even if $\sigma$ is finite but
large. Crucially for us, in the 
second and fifth steps, we have had to assume the absence of magnetic 
monopoles and magnetic currents. Hence magnetic helicity is conserved 
in a highly conducting plasma provided there are no magnetic monopoles. 
Then we might expect that if magnetic helicity is externally driven 
to large values in a plasma of extremely high electrical conductivity, 
magnetic monopoles might be produced to cause the helicity to dissipate. 

Our comments above also apply to the creation of sphalerons in the 
electroweak model since the sphaleron is precisely a bound state of an
electroweak monopole and antimonopole. The monopole pair in the sphaleron
are confined by a Z-string \cite{Achucarro:1999it} and cannot escape. Hence, 
if a sphaleron is produced, it will later decay. The production and decay of 
a sphaleron can lead to baryon number violation. So the discussion above 
suggests that it may be possible to observe baryon number violation by 
focusing magnetic helicity in a dense plasma\footnote{
As a first step in further investigations it may help to study sphaleron 
production in the 1+1 dimensional toy model proposed in 
Ref.~\cite{Mottola:1988ff}.}.

While we are still speculating, it is possible that the magnetic field
of certain atomic nuclei contain non-trivial magnetic helicity:
since magnetic helicity violates parity, the lightest stable candidate
nuclei are ${\rm Li}^7$ and ${\rm Be}^9$ \cite{nucleardata}. 
The scattering of a large number of such nuclei, $\sim 10^3$ based 
on energy requirements, seems like a possible path to concentrate 
magnetic helicity in a small spatial volume, perhaps sufficient to form 
a sphaleron. One way to investigate this 
process theoretically would be to use the description of nuclei in terms 
of Skyrmions. To study the feasibility of baryon number violation in the
collision of ``helical nuclei'', we would need to couple the Skyrmion fields 
to electroweak gauge fields, then consider the scattering of corresponding 
``helical Skyrmions'' in an attempt to find in-states that lead to sphaleron 
production, and thus baryon number violation. 

To conclude, we have studied the creation of solitons in a model that 
contains a twist degree of freedom in addition to the ones that are 
used for building the soliton. In our model in Sec.~\ref{O3model}, 
twist is an essential ingredient for the production of solitons 
because the dynamics in the zero twist sector is completely integrable.
We have explicitly shown a range of parameters that lead to soliton
creation. We have also discussed some applications of these ideas
to the production of vortices and monopoles, and baryon number violation
via the production of electroweak sphalerons.

\begin{acknowledgments}
I am grateful to Gil Speyer, Pascal Vaudrevange, and Amit Yadav 
for numerical help, and to Andrei Belitsky, Rich Lebed, and 
Juan Maldacena for discussions. The numerical work was done on 
the cluster at 
the ASU Advanced Computing Center. I also thank the Institute 
for Advanced Study for hospitality and the Department of Energy
for grant support at ASU.
\end{acknowledgments}

\appendix

\section{Numerical Notes}
\label{appa}

The most straightforward approach to solving the equations of motion is
to use spherical coordinates as in Eqs.~(\ref{thetaeq}), (\ref{phieq}). 
The issue there is that spherical coordinates are not globally well-defined 
and one must use two different patches, say, $(\theta_z,\phi_z)$ with respect 
to the $z-$axis and $(\theta_x,\phi_x)$ with respect to the $x-$axis. If 
$\theta_z$ becomes small, we can solve the evolution equation in terms of 
$\theta_x$ and $\phi_x$, and then transform to $\theta_z$ and $\phi_z$. 
We found that this scheme introduced unacceptable numerical errors in the 
boundary region of the two patches and we decided not to pursue this scheme.

Instead we decided to work with the vector representation (${\vec n}$) 
where the equation of motion can be written in first order form as
\begin{eqnarray}
\partial_t {\vec n} &=& {\vec m} \\
\partial_t {\vec m} &=& \partial_x^2{\vec n} 
          - [{\vec m}^2 - (\partial_x{\vec n})^2]{\vec n}
          - n_3 (n_3{\vec n}-{\vec e}_3)
\end{eqnarray}
We evolved these equations using the explicit second order Crank-Nicholson 
method with two iterations \cite{Teukolsky:1999rm}. We also implemented
boundary conditions, though in practice our lattice was large enough that
the boundaries were irrelevant.

The issue now is that when we discretize the above equations, they do not 
preserve the condition ${\vec n}^2 =1$ and the orthogonality condition 
${\vec n}\cdot {\vec m} =0$. To remedy this, we enforced these constraints
at every time step by rescaling ${\vec n}$ so as to get a unit vector, and
by subtracting out from ${\vec m}$ the component parallel to ${\hat n}$.
To our satisfaction, this procedure conserved energy at the percent level 
and the evolution also remained stable.


\begin{thebibliography}{99}


\bibitem{Mandelstam:1975hb}
  S.~Mandelstam,
  Phys.\ Rev.\  D {\bf 11}, 3026 (1975).

\bibitem{Dutta:2008jt}
  S.~Dutta, D.~A.~Steer and T.~Vachaspati,
  Phys.\ Rev.\ Lett.\  {\bf 101}, 121601 (2008)
  [arXiv:0803.0670 [hep-th]].

\bibitem{Romanczukiewicz:2010eg}
  T.~Romanczukiewicz and Y.~Shnir,
  Phys.\ Rev.\ Lett.\  {\bf 105}, 081601 (2010)
  [arXiv:1002.4484 [hep-th]].

\bibitem{Demidov:2011eu}
  S.~V.~Demidov and D.~G.~Levkov,
  JHEP {\bf 1106}, 016 (2011)
  [arXiv:1103.2133 [hep-th]].

\bibitem{Copi:2008he}
  C.~J.~Copi, F.~Ferrer, T.~Vachaspati, A.~Achucarro,
  Phys.\ Rev.\ Lett.\  {\bf 101}, 171302 (2008).
  [arXiv:0801.3653 [astro-ph]].

\bibitem{Chu:2011tx}
  Y.~-Z.~Chu, J.~B.~Dent, T.~Vachaspati,
  Phys.\ Rev.\ {\bf D83}, 123530 (2011).  
  [arXiv:1105.3744 [hep-th]].

\bibitem{Zamolodchikov:1978xm}
  A.~B.~Zamolodchikov and A.~B.~Zamolodchikov,
  Annals Phys.\  {\bf 120}, 253 (1979).

\bibitem{Dashen:1975hd}
  R.~F.~Dashen, B.~Hasslacher, A.~Neveu,
  Phys.\ Rev.\  {\bf D11}, 3424 (1975).

\bibitem{Nishida:2009}
M.~Nishida, Y.~Furukawa, T.~Fujii, N.~Hatakenaka,
Phys.\ Rev.\ {\bf E80}, 036603 (2009).

\bibitem{ongoingwork} 
T.~Vachaspati and P.~Vaudrevange, in progress.

\bibitem{Taubes:1982ie}
  C.~H.~Taubes,
  Commun.\ Math.\ Phys.\  {\bf 86}, 257 (1982).
  Commun.\ Math.\ Phys.\  {\bf 86}, 299 (1982).

\bibitem{Ruber:1985}
B.~R\"{u}ber, Ph.D. thesis, University of Bonn, 1985.

\bibitem{Kleihaus:1999sx}
  B.~Kleihaus, J.~Kunz,
  Phys.\ Rev.\  {\bf D61}, 025003 (2000).
  [hep-th/9909037].

\bibitem{Dashen:1974ck}
  R.~F.~Dashen, B.~Hasslacher, A.~Neveu,
  Phys.\ Rev.\  {\bf D10}, 4138 (1974).

\bibitem{Manton:1983nd}
  N.~S.~Manton,
  Phys.\ Rev.\  {\bf D28}, 2019 (1983).

\bibitem{Manton:2004tk}
  N.~S.~Manton, P.~Sutcliffe,
  ``Topological solitons,''
  Cambridge, UK: Univ. Pr. (2004) 493 p.

\bibitem{Kleihaus:1991ks}
  B.~Kleihaus, J.~Kunz, Y.~Brihaye,
  Phys.\ Lett.\  {\bf B273 } 100 (1991).

\bibitem{Kunz:1989}
  J.~Kunz and Y.~Brihaye,
  Phys.\ Lett.\  {\bf B216 } 353 (1989).

\bibitem{Yaffe:1989ms}
  L.~G.~Yaffe,
  Phys.\ Rev.\  D {\bf 40}, 3463 (1989).

\bibitem{Hellmund:1991ub}
  M.~Hellmund and J.~Kripfganz,
  Nucl.\ Phys.\  B {\bf 373}, 749 (1992).

\bibitem{Achucarro:1999it}
  A.~Achucarro, T.~Vachaspati,
  Phys.\ Rept.\  {\bf 327}, 347-426 (2000).
  [hep-ph/9904229].

\bibitem{Mottola:1988ff}
  E.~Mottola and A.~Wipf,
  Phys.\ Rev.\  D {\bf 39}, 588 (1989).

\bibitem{nucleardata}
http://www.nndc.bnl.gov/wallet/

\bibitem{Teukolsky:1999rm}
  S.~A.~Teukolsky,
  Phys.\ Rev.\  {\bf D61}, 087501 (2000).
  [gr-qc/9909026].

\end{thebibliography}
\end{document}